\documentclass[conference]{IEEEtran}
\IEEEoverridecommandlockouts

\usepackage{cite}
\usepackage{amsmath,amssymb,amsfonts}
\usepackage{algorithmic}
\usepackage{graphicx}
\usepackage{textcomp}
\usepackage{xcolor}
\usepackage[hidelinks]{hyperref}
\usepackage{listings}
\lstdefinestyle{json}{
  basicstyle=\ttfamily\footnotesize,
  breaklines=true,
  frame=single,
  columns=fullflexible,
  showstringspaces=false,
}
\def\BibTeX{{\rm B\kern-.05em{\sc i\kern-.025em b}\kern-.08em
    T\kern-.1667em\lower.7ex\hbox{E}\kern-.125emX}}
\begin{document}

\title{Catching UX Flaws in Code: Leveraging LLMs to Identify Usability Flaws at the Development Stage\\
}

\author{\IEEEauthorblockN{Nolan Platt}
\IEEEauthorblockA{
\textit{Virginia Tech}\\
Blacksburg, Virginia \\
nolanplatt@vt.edu}
\and
\IEEEauthorblockN{Ethan Luchs}
\IEEEauthorblockA{
\textit{Virginia Tech}\\
Blacksburg, Virginia\\
ethanluchs@vt.edu}
\and
\IEEEauthorblockN{Sehrish Nizamani}
\IEEEauthorblockA{
\textit{Virginia Tech}\\
Blacksburg, Virginia\\
sehrishbasir@vt.edu}
}

\maketitle

\begin{abstract}
Usability evaluations are essential for ensuring that modern interfaces meet user needs, yet traditional heuristic evaluations by human experts can be time‑consuming and subjective, especially early in development. This paper investigates whether large language models (LLMs) can provide reliable and consistent heuristic assessments at the development stage. By applying Jakob Nielsen's ten usability heuristics to thirty open‑source websites, we generated over 850 heuristic evaluations in three independent evaluations per site using a pipeline of OpenAI's GPT-4o. For issue detection, the model demonstrated moderate consistency, with an average pairwise Cohen's Kappa of 0.50 and an exact agreement of 84\%. Severity judgments showed more variability: weighted Cohen's Kappa averaged 0.63, but exact agreement was just 56\%, and Krippendorff's Alpha was near zero. These results suggest that while GPT-4o can produce internally consistent evaluations, especially for identifying the presence of usability issues, its ability to judge severity varies and requires human oversight in practice. Our findings highlight the feasibility and limitations of using LLMs for early-stage, automated usability testing, and offer a foundation for improving consistency in automated User Experience (UX) evaluation. To the best of our knowledge, our work provides one of the first quantitative inter‑rater reliability analyses of automated heuristic evaluation and highlights methods for improving model consistency.
\end{abstract}


\section{Introduction}
Usability is a critical quality attribute for web applications, directly influencing user satisfaction, adoption, and long-term success~\cite{b1}. A feature-rich site still fails if users struggle to accomplish basic tasks~\cite{b2}, leading to costly redesigns and negative feedback. Detecting and resolving such usability flaws as early as possible is therefore highly desirable. Traditionally, usability issues are discovered through \textit{user testing}, where real users are observed while completing tasks~\cite{b2}, and via \textit{expert heuristic evaluations}, where specialists review the interface against established principles. While effective, these approaches demand significant time, expertise, and coordination~\cite{b4, b5}. Small teams or early-stage projects often lack the resources for comprehensive usability studies, allowing interface problems to remain hidden until after deployment~\cite{b6}. This resource gap motivates our investigation into automated usability assessments~\cite{b7} that can be run continuously during development with minimal human effort.

Recent advances in Large Language Models (LLMs), especially with regard to Natural Language Processing (NLP), offer a novel approach to automate heuristic evaluations. LLMs are deep neural network models trained on vast text corpora and advanced methods in NLP; in some cases, these models are trained to interpret and understand code and images, enabling them to understand and generate complex language and reason about structured information. The past few years have seen a rapid development in LLMs ability to perform software engineering tasks, with specific regard to code generation, code review, and bug detection~\cite{b8}, and with some models able to simultaneously process files, interpret the source code, and understand natural language instructions, the applications of LLMs are nearly endless~\cite{b9}. This raises the intriguing possibility that an LLM could serve as an \textit{automated usability evaluator}, essentially acting as a virtual ``expert'' that inspects a web application's interface and code for usability guideline violations. If viable, such an approach could rapidly flag UX issues early on, providing feedback to developers without the need for costly usability studies.

This research investigates whether state-of-the-art LLMs can reliably identify usability flaws in web applications during early development by analyzing the respective source code. In particular, we examine the use of OpenAI GPT-4o to perform heuristic evaluations on web interfaces. We leverage Jakob Nielsen's ten usability heuristics~\cite{b4} as a structured framework for analysis, prompting the model to check each heuristic with respect to potential issues in the given website's implementation. To evaluate consistency and reliability, we conduct an experiment on \textit{30 open-source web applications}, instructing the model to review each application \textit{three separate times} in fresh sessions. By treating each independent LLM session as a rater, we then compute inter-rater reliability metrics through various statistical measures: \textbf{Cohen's Kappa}, \textbf{Fleiss's Kappa}, and \textbf{Krippendorff's Alpha}. Through these measures, we effectively  quantify how consistent the model identifies the same issue and severity rating across independent sessions. High agreement would indicate that the model's usability evaluations are stable and repeatable, whereas low agreement would signal variability or randomness in its responses. Through this study, we aim to assess the practical feasibility of LLM-driven usability inspection and understand its current strengths and limitations.

\section*{Key Contributions}

In summary, the key contributions of this work are as follows:

\textbf{1. LLM-Based Heuristic Evaluation Methodology.}
We propose a novel approach to automated usability testing using a large language model. Our method translates Nielsen's usability principles into structured prompts that guide GPT-4o to inspect web application source code for usability concerns, requiring no domain-specific fine-tuning of the model. We leverage a customized pipeline of GPT-4o that standardizes prompting and outputs results in JavaScript Object Notation (JSON) format.

\textbf{2. Empirical Study on Web Applications.}
We conduct a comprehensive evaluation on 30 open-source websites, applying the LLM-driven heuristic evaluation to each. This design allows us to observe the types of usability issues a model can detect in real-world web projects and to gather qualitative examples of its feedback.

\textbf{3. Inter-Rater Reliability Analysis.}
To our knowledge, we are among the first to examine the consistency of an LLM's UX evaluations. By repeating the analysis in triplicate for each site and calculating Cohen's Kappa, Fleiss's Kappa, and Krippendorff's Alpha, we provide quantitative insight into the reliability of GPT-4o's judgments.

\textbf{4. Insights on LLM Efficacy for Usability Testing.}
We discuss the implications of our findings for the role of AI in usability testing. Our results shed light on the types of usability problems the model excels at or misses, and how it might be best integrated into development workflows.

The remainder of this paper is structured as follows: \textbf{\ref{background}} provides background on usability evaluation methods and reviews related work in AI-assisted usability assessment. \textbf{\ref{methodology}} then details our methodology for LLM-driven heuristic evaluation, including prompt design and experimental setup. \textbf{\ref{results}} presents the results of our study, focusing on the inter-rater reliability measures and examples of identified issues. \textbf{\ref{discussion}} offers a discussion on the findings, their implications, and limitations. Finally, \textbf{\ref{conclusion}} concludes the paper with a summary and outlook on future developments in automated usability testing.

\section{Background} \label{background}

\subsection{Usability Evaluation and Heuristic Methods}
Usability refers to how easy and effective it is for users to achieve their goals using a system. Key aspects include learnability, efficiency, error reduction, and user satisfaction. Common evaluation methods include user testing and heuristic evaluation. User testing involves observing participants complete tasks on an interface, offering rich empirical feedback but at the cost of time and resource investment. On the other hand, heuristic evaluation, introduced by Nielsen and Molich, requires experts to review an interface against recognized usability principles~\cite{b4, b5}. Nielsen's ten usability heuristics remain a standard, covering principles such as visibility of system status, user control, consistency, and error prevention.

Heuristic evaluations are faster and less resource-intensive than user testing but suffer from subjectivity and limited coverage. Different experts may identify different issues or disagree on severity, leading to inconsistent results~\cite{b5}. These inconsistencies are especially problematic in early-stage development, where iterative decisions depend on timely and actionable feedback. Moreover, many small teams lack access to usability experts, and even with multiple reviewers, variability persists. These challenges have motivated growing interest in automated heuristic evaluation as a way to reduce cost, improve coverage, and support continuous integration in agile workflows.

\subsection{AI-Powered Usability Assessment with LLMs}

To address the limitations of manual and expert-driven reviews, several tools have emerged to automate parts of the UX evaluation process. Widely used tools apply static rule-based analysis to detect violations of accessibility and usability standards in HTML and CSS. These systems are highly effective for identifying technical flaws—such as missing alternative text, improper contrast ratios, or non-semantic markup—but are fundamentally constrained by their deterministic nature. They operate by checking for predefined conditions, and as a result, they miss more nuanced, context-dependent problems such as ambiguous labels, ineffective navigation, or tone inconsistency.

Large Language Models (LLMs) offer a promising alternative to these hard-coded approaches. LLMs provide a semantic layer of reasoning that allows them to analyze both the structure and intent of an interface~\cite{b10}. Language models can be prompted with heuristic criteria and interpret code - including HTML, CSS, and JavaScript - to identify potential usability concerns~\cite{b11}. This enables such models to detect flaws that go beyond syntax, such as misleading User Interface (UI) metaphors, unclear instructions, or a lack of user control mechanisms. This flexibility positions LLMs as general-purpose evaluators capable of both qualitative and structured analysis. Recent work like UX-LLM~\cite{b12} uses multimodal approaches but requires rendered UIs, limiting early-stage applicability~\cite{b13}.

Our approach focuses on source code-only evaluation, leveraging LLMs to reason about usability directly from markup and scripting logic. This allows assessments to be conducted without the need for rendering or prototyping, enabling integration into early-stage development workflows. By encoding usability heuristics into structured prompts, we produce evaluations that include both quantitative severity ratings and detailed qualitative feedback.

To the best of our knowledge, prior work has not examined the reproducibility of such LLM-driven evaluations. Some studies report on detection accuracy or illustrative output, but few if any quantify how consistent these models are when applied multiple times under the same conditions. In human-centered design, inter-rater reliability (IRR) is a critical measure of evaluation quality—used to assess whether different evaluators (or the same evaluator across time) produce similar judgments. In this work, we adapt IRR metrics, including Cohen's Kappa, Fleiss's Kappa, and Krippendorff's Alpha, to assess agreement across multiple independent GPT-4o sessions.  

\section{Methodology} \label{methodology}

\subsection{Initial Approach} \label{sec:initialApproach}

Initial research centered around using a consistent and exact prompt across various usability evaluations. This led to flawed and - in some cases -  completely inaccurate results. GPT-4o initially struggled with consistently filling out templates, both in chat (i.e., directly to the user) and in downloadable document files. We quickly realized that we needed to design an acutely specific prompt, consisting of prerequisite knowledge, instructions for document processing, and clear guidelines on how output should be displayed. In natural language processing (NLP), there has been a significant challenge in fine-tuning models to comprehend the same instruction over time. Specifically, GPT-4o is known to behave inconsistently for semantically identical queries in different sessions. This caused significant issues in our initial data collection, in that the evaluations were frequently recorded wrong or the same heuristic was duplicated dozens of times.

\subsection{Data Collection} \label{dataCollection}
To address the underlying issues discussed in \ref{sec:initialApproach}, we designed a pipelined version of GPT-4o. OpenAI allows users to create their own "versions" of different LLM models, in that the session will already contain prerequisite knowledge, instructions, and address misconceptions that arise in NLP. With regard to the issues we faced, we were able to design the application to:
\begin{itemize}
    \item understand the user is uploading a compressed file containing a website's source code.
    \item have exact knowledge of Nielsen's heuristic principles and how they apply to modern web applications.
    \item only output the evaluation in a consistent JSON format.
\end{itemize}

See Appendix\ref{app:prompt} for the complete format and structure of the prompt. The prompt ensured JSON-only output for each of Nielsen's 10 heuristics. Through this, we were able to address the NLP challenges as best we could without domain-specific fine-tuning. We used three independent GPT-4o sessions per project to evaluate intra-model consistency and account for session-level prompt variability. The current methodology of conducting an evaluation is as follows:

\begin{enumerate}
    \item identify an open-source web application via GitHub or other version control platforms.
    \item verify the repository's licensing and our ability to use the code.
    \item download the repository as a compressed ZIP file.
    \item open three distinct sessions of the customized GPT.
    \item upload the same compressed file to each of the three sessions and download each resulting JSON file. 
\end{enumerate}

Projects were selected from public GitHub repositories under permissive licenses, focusing on early-stage front-end web applications (HTML, CSS, JavaScript only). After step five, each JSON file was then organized in the archived dataset~\cite{b16} at \verb|\results\<repoName>\<eval#>|, where:\newline
 \verb|repoName| represents the name of the repository.\newline
 \verb|eval#| represents the evaluation in ascending order: \verb|eval1|, \verb|eval2|, \verb|eval3|.\newline  By using three distinct evaluations for each website, we are able to quantify how GPT-4o performs with regard to consistency and accuracy over time. This consistent methodology proved highly effective, with GPT-4o consistently providing correctly structured JSON files, and not misinterpreting the natural language instructions.

 \section{Results} \label{results}
 To obtain accurate results from our study, we created an open-source repository that includes all results (in JSON) from  GPT-4o, along with the analysis code in a Jupyter notebook.
 This repository is publicly available at \href{https://github.com/nolanplatt/LLMUsabilityEvaluations}{\texttt{github.com/nolanplatt/LLMUsabilityEvaluations}}\cite{b16}. In this paper, all references to directories refer to this repository's root.
 
 \subsection{Issue Frequency Across Evaluations}

 Our analysis begins in \verb|\analysis.ipynb|, where we iterate through every folder in \verb|\results|, each representing an open-source website and the corresponding evaluations. For each site, we load the triplicate LLM evaluations into rows of a \verb|DataFrame|, with each row holding three independent evaluations corresponding to the same site. 

\begin{figure}[ht]
    \centering
    \includegraphics[width=0.8\linewidth]{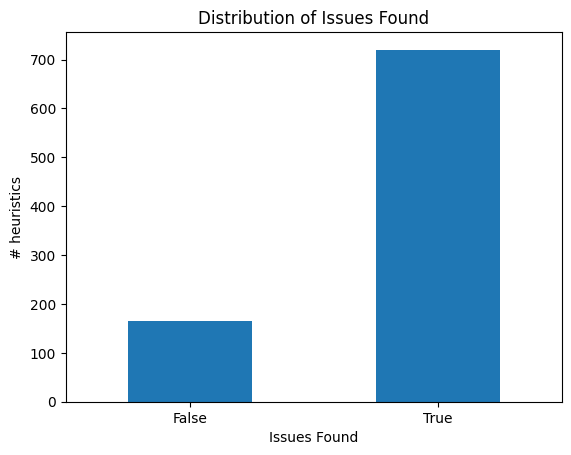}
    \caption{Distribution of issues found across all evaluations. A large majority of heuristics were flagged as having usability issues by the model.}
    \label{fig:issueGraph}
\end{figure}

To visualize how often GPT-4o flags usability issues, Fig.~\ref{fig:issueGraph} shows the percentage of heuristics marked as having an issue across all 30 websites. Each heuristic was reviewed three times per site, totaling 900 evaluations. The overwhelming majority of evaluations identify at least one usability flaw, reflecting the fact that our dataset largely consists of early-stage or demo-level web projects.

This high detection rate demonstrates that GPT-4o frequently identifies flaws when reviewing source code with respect to Nielsen's heuristics. Nonetheless, frequent detection does not necessarily imply consistent or accurate evaluation. Rather, this indicates a tendency toward sensitivity, often flagging something as an issue, which we further examine through severity analysis.

\subsection{Distribution of Severity Ratings}\label{sec:distribution}
To explore how serious GPT-4o identified each issue to be, we analyzed severity scores across all sessions and sites. Table~\ref{tab:severity} summarizes the full distribution. Ratings range from 0 (no issue) to 4 (usability catastrophe).

\begin{table}[htbp] 
  \centering
  \caption{Distribution of Severity Ratings}
  \label{tab:severity}
  \begin{tabular}{|c|c|}
    \hline
    \textbf{Severity Level} & \textbf{Count} \\
    \hline
    0 (No Issue) & 164 \\
    1 (Cosmetic) & 251 \\
    2 (Minor)    & 306 \\
    3 (Major)    & 155 \\
    4 (Catastrophic) & 9 \\
    \hline
  \end{tabular}
\end{table}

This distribution highlights the model's tendency to flag minor and moderate issues far more frequently than critical failures. The total number of severity ratings shown in Table~\ref{tab:severity} is $885$. While 900 heuristic evaluations were expected (30 websites $\times$ 10 heuristics $\times$ 3 sessions), 15 entries were excluded due to missing or malformed severity values. These incomplete outputs were dropped during preprocessing to ensure consistency in statistical comparisons.

\subsection{Agreement on Issue Presence}\label{sec:issuePresence}
We first examined whether GPT-4o agreed on whether a usability issue \textit{existed}. For this, we calculated \textbf{Cohen's Kappa} across each pair of evaluations for every site, as represented by Equation~\eqref{cohenKappa}.

\begin{equation}
\kappa = \frac{p_o - p_e}{1 - p_e}\label{cohenKappa}
\end{equation}
where \(p_o\) is the observed agreement and \(p_e\) is the expected agreement by chance. This was applied to the evaluation pairs for each site, as shown in Table~\ref{tab:cohenBinary}.

  Since the result is binary, weighting is unnecessary. The $\kappa$ values in all comparisons are moderately consistent, ranging from 0.47 to 0.53, with exact agreement rates between 82.3\% and 85.0\%. This suggests that GPT-4o is moderately reliable in consistently identifying whether or not a heuristic violation exists, even in multiple, distinct trials.

  The high agreement percentages (\textgreater82\%) show that the model does not flag issues randomly, but follows repeatable logic to determine when a heuristic has been violated or not.
\begin{table}[htbp]
    \centering
    \caption{Unweighted Cohen's Kappa Across All Issue Pairs}
    \label{tab:cohenBinary}
    \begin{tabular}{|c|c|c|c|}
        \hline
        \textbf{Comparison} & \textbf{Cohen's $\kappa$} & \textbf{Agreement} & \textbf{Exact Percent} \\ \hline
        eval1-eval2 & 0.471 & 247/300 & 82.3 \\ 
        eval1-eval3 & 0.530 & 255/300 & 85.0 \\ 
        eval2-eval3 & 0.502 & 254/300 & 84.7 \\ \hline
    \end{tabular}

\end{table}

  We can further evaluate the significance of this data using multi-rater agreement metrics, including: 

\noindent\textbf{Fleiss's Kappa.}
Used to assess agreement among three raters. The general formula utilized is expressed in Equation~\eqref{fleissKappa}
.\begin{equation}
\kappa = \frac{\bar{P} - \bar{P_e}}{1 - \bar{P_e}}\label{fleissKappa}
\end{equation}
where \(\bar{P}\) is the mean of observed agreements, and \(\bar{P_e}\) is the mean of expected agreement.

\noindent\textbf{Krippendorff's Alpha.}
A general measure that accommodates missing data and all data types, as shown in Equation~\eqref{krippendorff}.
\begin{equation}
\alpha = 1 - \frac{D_o}{D_e}\label{krippendorff}
\end{equation}
where \(D_o\) is the observed disagreement and \(D_e\) is the expected disagreement.

Both \eqref{fleissKappa} and \eqref{krippendorff}  were applied to the same data for issue pairs, with the results shown in Table~\ref{tab:multiRater}.

While Cohen's and Fleiss's Kappa metrics suggest moderate agreement across different sessions (e.g., average $\kappa$ $\approx$ 0.5), Krippendorff's Alpha returned a value slightly negative. This is not a calculation error: it reflects how the metric is conservative with respect to class imbalance and small differences between sessions.

\begin{table}[htbp]
    \centering
     \caption{Multi-Rater Agreement Metrics}
    \label{tab:multiRater}
    \begin{tabular}{|c|c|}
        \hline
        \textbf{Metric} & \textbf{Value} \\ \hline
        Krippendorff's $\alpha$ & -0.000234 \\ 
        Fleiss's $\kappa$       & 0.500  \\ \hline
    \end{tabular}
   
\end{table}
\subsection{Agreement on Severity Ratings}

\begin{figure}[htbp]
    \centering  
    \includegraphics[width=1\linewidth]{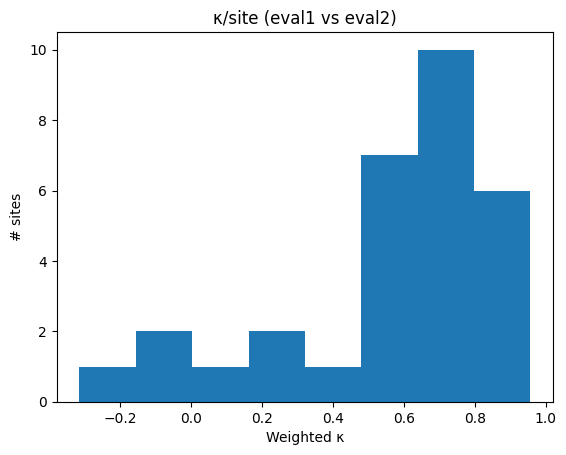}
    \caption{Distribution of weighted Cohen's Kappa values per site (eval1 vs eval2) for severity ratings.}
    \label{fig:irrGraph}
\end{figure}

As described in \ref{sec:distribution}, severity ratings range from 0 (no issue) to 4 (usability catastrophe). We are thus looking at five possible ratings the LLM can rate \textit{each heuristic}. So, statistically comparing these results cannot be achieved using the binary approach in \ref{sec:issuePresence}. We must instead utilize a weighted Cohen's Kappa~\cite{b17}. Fig.~\ref{fig:irrGraph} represents the data-wide distribution of Cohen's Kappa for \verb|eval1 vs eval2|. As visualized, most sites achieved \(\kappa >= 0.6\), indicating substantial agreement, though a few had lower or even negative agreement, typically due to imbalanced detection rates between the two.
Because severity is ordinal, we can understand the agreements in the data through a \textbf{Weighted Cohen's Kappa}. Equation~\eqref{cohenSeverity} applies penalties to disagreements based on the significance of the displacement between two ratings.

\begin{equation}
\kappa_w = 1 - \frac{\sum w_{i,j} o_{i,j}}{\sum w_{i,j} e_{i,j}}\label{cohenSeverity}
\end{equation}
where \(w_{i,j}\) is a weight matrix and \(o_{i,j}, e_{i,j}\) are observed counts.
The weighted $\kappa_w$ was then computed for all severity pairs, as shown in Table~\ref{tab:allWeighted}.

GPT-4o is not only consistent in identifying whether a usability issue exists, but also exhibits reasonable agreement in how severe those issues are. Severity judgments do show more variability than binary issue presence, but this is expected; the severity data is ordinal, rather than binary. Most sessions still landed within a close scoring range. The weighted $\kappa$ values near or above 0.6 (see Table~\ref{tab:allWeighted}) imply moderate agreement between evaluations, reinforcing the notion that GPT-4o is capable of reliably ranking issue significance.

In particular, this consistency is crucial for developers or small teams in prioritizing fixes and, thus, workflow. Even if the exact severity scores vary slightly, the model does demonstrate a consistent understanding of what constitutes a minor usability flaw compared to a catastrophic issue. However, it is vital to understand that this is \textit{internal consistency}, and does not imply actual correctness. The model's ratings should be used to guide - not replace - expert judgment. 

\begin{table}[htbp]
    \centering
    \caption{Weighted Cohen's Kappa Across All Severity Pairs}
    \label{tab:allWeighted}
    \begin{tabular}{|c|c|c|c|}
        \hline
        \textbf{Comparison} & \textbf{$\kappa_w$} & \textbf{Agreement} & \textbf{Exact Percent}\\ \hline
        eval1-eval2   & 0.625361	   & 169/300  & 56.3  \\ 
        eval1-eval3   & 0.668431	   & 171/300	   & 57  \\ 
        eval2-eval3  & 0.596722	 &167/300  & 55.6  \\ \hline
    \end{tabular}
\end{table}

Representative JSON excerpts are provided in Appendix\ref{app:example1} and Appendix\ref{app:example2} to illustrate typical outputs for two distinct heuristics; the full dataset is available via the Figshare archive~\cite{b16}.

When examining individual heuristics (issue presence, unweighted), agreement was highest for \textit{Error Prevention} ($\kappa=0.42$--$0.55$, mean $0.47$) and \textit{Match System and Real World} ($\kappa=0.27$--$0.57$, mean $0.40$), and lowest for \textit{Help and Documentation} ($\kappa=-0.07$--$0.19$, mean $0.08$) and \textit{Consistency and Standards} ($\kappa=-0.03$--$0.30$, mean $0.08$). Complete per-heuristic values are in Appendix\ref{app:perheuristic}. Each heuristic used 30 sites $\times$ 3 sessions; incomplete entries were dropped as explained in Section~\ref{sec:distribution}.

\section{Discussion} \label{discussion}
Our findings show that automated heuristic evaluations can produce consistent results, especially when identifying whether a usability issue exists. The relatively high agreement percentages (82–85\%) and moderate Cohen's Kappa values ($\approx$0.5) show that GPT-4o applies its internal logic consistently across independent sessions. This is a promising result, particularly for teams lacking access to UX professionals.

Severity evaluations showed agreement rates of 56-57\% with weighted Cohen's Kappa near 0.6. Although not as strong as the binary classification, this still indicates inter-rater reliability, even in the ordinal nature of the severity ratings. More remarkably, this level of reliability is achieved without domain-specific fine-tuning. In traditional human evaluations, severity judgments are also known to vary, even among expert raters~\cite{b2}. The consistency observed here suggests that GPT-4o may offer comparable approaches when prioritizing usability problems (and ratings).

Krippendorff's Alpha value being slightly negative shows how small inconsistencies across sessions are \textit{heavily} penalized by this conservative metric. This reflects the metric's sensitivity to systematic disagreement, not model failure.

Our finding shows that LLMs can serve as \textit{preliminary} tools for automated usability evaluations, especially when expert resources are limited. The consistency shown does not imply correctness. Future work should compare model output to expert evaluations.

\section{Conclusion} \label{conclusion}

In this paper, we demonstrate the potential of large language models to perform consistent heuristic evaluations on website source code. We find that GPT-4o exhibits moderate consistency, particularly for issue detection. Severity ratings show reasonable internal consistency with slight variation, highlighting the model's ability to distinguish between different levels of usability concern.

Although LLMs are not a replacement for expert evaluations, our results suggest they can serve as consistent early-stage tools to identify usability flaws. This capability is especially valuable for small teams where traditional evaluations may be impractical or time consuming. Our methodology lays the foundation for reproducible AI-assisted usability evaluations.

Future work may explore comparing model output with expert human evaluations, improving prompt robustness, and domain-specific fine-tuning. As LLMs continue to improve in accuracy and natural language processing, their role in usability evaluations will likely expand from supporting human evaluators to driving entirely new systems of automated usability assessments.

\section*{Acknowledgment} 
We would like to thank the \href{https://hci.icat.vt.edu/}{Virginia Tech Center for Human-Computer Interaction (CHCI)} for providing invaluable feedback during the preliminary stages of our research.

\appendices
\section*{Appendix}

\subsection{LLM Prompt Instructions} \label{app:prompt}

As discussed, we utilized a customized and consistent version of GPT-4o. This ensured the model received identical instructions and context across evaluations. The prompt used is shown below.

\begin{quote}
\ttfamily
You are a usability evaluation expert specializing in website usability based on code analysis. Users will provide a .zip file containing the full code of a website (HyperText Markup Language (HTML), Cascading Style Sheets (CSS), JavaScript (JS)). Your task is to perform a detailed heuristic evaluation based on Jakob Nielsen's 10 usability heuristics. For each heuristic, you will: \\
- Assign a SeverityRating (0 through 4) \\
- Indicate IssueFound (true/false) \\
- Write an IssueDescription \\
- Provide CodeReference (file name(s) and line number(s)) \\
- Provide a CodeSnippet \\
- Complete EvaluationAnswers (with explanations) \\
- Offer a clear Recommendation \\

If no issue is found, SeverityRating must be 0, IssueFound must be false, and placeholders used as directed. Always return a single, clean, valid JSON object for all 10 heuristics, without skipping any heuristic or reusing responses. Use professional, detailed language. Do not proceed unless the .ZIP is provided. Verify internally that all heuristics are evaluated and JSON is valid before replying. \\

 Accuracy and clarity are mandatory. Nothing can be skipped or overlooked. \\

Always send the output as a downloadable .JSON file. The format must always be:
\begin{lstlisting}[style=json]
[
  {
    "Heuristic": "Heuristic",
    "SeverityRating": 0-4,
    "IssueFound": true/false,
    "IssueDescription": "Issue Description",
    "CodeReference": "Code reference",
    "CodeSnippet": "Code snippet",
    "EvaluationAnswers": { ... },
    "Recommendation": "Clear and actionable recommendation."
  }
]
\end{lstlisting}
\end{quote}

\vspace{12pt}
\newpage
\subsection{Example JSON Excerpt: Visibility of System Status} \label{app:example1}
\lstset{style=json}
\begin{lstlisting}
{
  "Heuristic": "Visibility of system status",
  "SeverityRating": 3,
  "IssueFound": true,
  "IssueDescription": "The system lacks loading indicators and confirmation messages, which leaves users unsure whether actions such as search or navigation were successful.",
  "CodeReference": "blank.html: Line 150-160",
  "CodeSnippet": "<form class=\"d-none d-sm-inline-block form-inline mr-auto ml-md-3 my-2 my-md-0 mw-100 navbar-search\">...</form>",
  "Recommendation": "Add visual loading indicators, confirmation messages, and inline validation to inform users about system status clearly and promptly."

}
\end{lstlisting}
\newpage
\subsection{Example JSON Excerpt: Help and Documentation} \label{app:example2}
\lstset{style=json}
\begin{lstlisting}
{
  "Heuristic": "Help and documentation",
  "SeverityRating": 2,
  "IssueFound": true,
  "IssueDescription": "No embedded help or documentation is available in the UI for guidance.",
  "CodeReference": "All HTML templates"
  "CodeSnippet": "<body>...</body>",
  "Recommendation": "Provide inline help icons, tooltips, or links to external documentation for user assistance."
}
\end{lstlisting}

\subsection{Per-Heuristic Kappa Values} \label{app:perheuristic}
\begin{table}[htbp]
    \centering
    \caption{Cohen's Kappa Values by Heuristic Across Evaluation Pairs}
    \label{tab:perheuristic}
    \resizebox{\columnwidth}{!}{%
    \begin{tabular}{|l|c|c|c|}
    \hline
    \textbf{Heuristic} & \textbf{$\kappa$ (1-2)} & \textbf{$\kappa$ (1-3)} & \textbf{$\kappa$ (2-3)} \\
    \hline
    Aesthetic And Minimalist Design & 0.328 & 0.143 & 0.154 \\
    Consistency And Standards & -0.024 & 0.302 & -0.029 \\
    Error Prevention & 0.416 & 0.551 & 0.438 \\
    Flexibility And Efficiency Of Use & 0.103 & 0.280 & 0.032 \\
    Help And Documentation & -0.068 & 0.123 & 0.188 \\
    Help Users Recognize Errors & 0.406 & 0.292 & 0.231 \\
    Match System And Real World & 0.569 & 0.364 & 0.267 \\
    Recognition Rather Than Recall & 0.238 & 0.211 & 0.132 \\
    User Control And Freedom & 0.327 & 0.534 & -0.010 \\
    Visibility Of System Status & 0.427 & 0.298 & 0.252 \\
    \hline
    \end{tabular}%
    }
\end{table}

\end{document}